%% file: main.tex
\documentclass[conference]{IEEEtran}
\IEEEoverridecommandlockouts

\usepackage{cite}
\usepackage{amsmath,amssymb,amsfonts}
\usepackage{algorithmic}
\usepackage{graphicx}
\usepackage{textcomp}
\usepackage{xcolor}
\usepackage{caption}
\usepackage{subcaption}
\usepackage{hyperref}
\usepackage[utf8]{inputenc}
\usepackage{varwidth}
\usepackage{tabularx}
\usepackage{physics}
\usepackage{enumitem}

\newcommand{\figref}[1]{Figure~\ref{#1}}
\newcommand{\secref}[1]{Section~\ref{#1}}

\newcommand{\tabref}[1]{Table~\ref{#1}}

\def\BibTeX{{\rm B\kern-.05em{\sc i\kern-.025em b}\kern-.08em
    T\kern-.1667em\lower.7ex\hbox{E}\kern-.125emX}}

\begin{document}

\title{{\huge Enhancing LLM-based Quantum Code Generation with Multi-Agent Optimization and Quantum Error Correction}\\
}
\author{\IEEEauthorblockN{Charlie Campbell}
\IEEEauthorblockA{\textit{Department of Computing} \\
\textit{Imperial College London}\\
London, UK \\
charlie.campbell22@imperial.ac.uk}
\and
\IEEEauthorblockN{Hao (Mark) Chen}
\IEEEauthorblockA{\textit{Department of Computing} \\
\textit{Imperial College London}\\
London, UK \\
hao.chen20@imperial.ac.uk}
\and
\IEEEauthorblockN{Wayne Luk}
\IEEEauthorblockA{\textit{Department of Computing} \\
\textit{Imperial College London}\\
London, UK \\
w.luk@imperial.ac.uk}
\and
\IEEEauthorblockN{Hongxiang Fan}
\IEEEauthorblockA{\textit{Department of Computing} \\
\textit{Imperial College London}\\
London, UK \\
hongxiang.fan@imperial.ac.uk}
}

\maketitle

\begin{abstract}

Multi-agent frameworks with Large Language Models (LLMs) have become promising tools for generating general-purpose programming languages using test-driven development, allowing developers to create more accurate and robust code. 
However, their potential has not been fully unleashed for domain-specific programming languages, where specific domain exhibits unique optimization opportunities for customized improvement.
In this paper,
we take the first step in exploring multi-agent code generation for quantum programs.
By identifying the unique optimizations in quantum designs such as quantum error correction,
we introduce a novel multi-agent framework tailored to generating accurate, fault-tolerant quantum code.
Each agent in the framework focuses on distinct optimizations, iteratively refining the code using a semantic analyzer with multi-pass inference, alongside an error correction code decoder.
We also examine the effectiveness of inference-time techniques, like Chain-of-Thought (CoT) and Retrieval-Augmented Generation (RAG) in the context of quantum programming, uncovering observations that are different from general-purpose code generation.
To evaluate our approach, we develop a test suite to measure the impact each optimization has on the accuracy of the generated code. 
Our findings indicate that techniques such as structured CoT significantly improve the generation of quantum algorithms by up to 50\%. 
In contrast, we have also found that certain techniques such as RAG show limited improvement, yielding an accuracy increase of only 4\%.
Moreover, we showcase examples of AI-assisted quantum error prediction and correction, demonstrating the effectiveness of our multi-agent framework in reducing the quantum errors of generated quantum programs.
\end{abstract}


\input{1_introduction}
\input{2_background}
\input{3_framework}
\input{4_optimization}

\input{5_experiments}
\input{6_conclusion}

\section*{Acknowledgment}

The support of UK EPSRC (grant number EP/W032635/1,
EP/V028251/1, EP/S030069/1 and EP/X036006/1), Intel and AMD is gratefully acknowledged.


\newpage

\bibliographystyle{IEEEtran}
\bibliography{0_ref}

\end{document}

%% file: 1_introduction.tex
\section{Introduction}

Recent advances in Large Language Models (LLMs) have revolutionized a wide range of AI applications, including natural language understanding, machine translation, and automated content generation~\cite{hadi2024large}. 
Building on the exceptional natural language processing capabilities of LLMs, recent research has extended into the exploration of multi-agent frameworks, where each agent is driven by an LLM to facilitate complex interactions in complicated environments.
These multi-agent frameworks form the foundation for embodied AI systems~\cite{duan2022survey}, paving the way for a future where AI plays a critical role in assisting and supporting daily human activities.

A key application of LLM is in automatic software development, where AI can assist developers in generating increasingly high-quality code~\cite{jiang2024survey}. 
Various LLMs tailored for code generation, such as StarCoder~\cite{lozhkov2024starcoder} or CodeLlama~\cite{roziere2023code}, have been introduced to produce high-quality code and test suites. 
Furthermore, recent research has explored multi-agent frameworks, such as AgentCoder~\cite{huang2023agentcoder}, to enhance collaborative code generation.
However, most of these efforts have focused on general-purpose programming languages like C++ and Python, with relatively less emphasis on domain-specific programming languages.

As a pioneering force in quantum computing, IBM has advanced the application of LLMs for quantum program generation, demonstrating the potential of LLMs in producing domain-specific programs~\cite{dupuis2024qiskit}.
To facilitate the development of LLM in quantum computing,
they introduce training datasets, a customized LLM and testbench for quantum program generation.
While this work lays a solid foundation for leveraging LLMs in quantum computing, several key challenges remain in optimizing LLM-assisted quantum code generation:

\begin{itemize}
    \item The rapid pace of quantum computing advancements requires frequent updates to code libraries, resulting in a lack of high-quality, up-to-date training data. Models trained on data that are only a few months old can quickly become obsolete due to the release of new libraries or algorithms.
    \item Quantum programs generated by LLMs, with only standard supervised fine-tuning, tend to be error-prone. It is essential to explore the role of advanced prompt engineering in improving the reliability of LLM-assisted quantum code generation.
    \item Prior approaches have applied traditional code generation techniques used for general-purpose programming languages to quantum computing without accounting for domain-specific optimizations, such as Quantum Error Correction (QEC). This results in suboptimal performance. For instance, to meet the stringent standards of quantum code generation, LLMs must be capable of producing fault-tolerant quantum code using QEC techniques.
\end{itemize}

To address the aforementioned challenges,
this paper proposes a novel multi-agent quantum code generation framework that integrates iterative multi-pass optimization and automatic quantum error correction.
Inspired by OpenAI's \textit{GPT-o1}~\footnote{\url{https://openai.com/index/introducing-openai-o1-preview/}} which emphasizes inference time optimization over training time effort, we demonstrate that multiple inference-time optimization techniques can effectively mitigate the issue introduced by limited training data. 
To enhance the efficiency of quantum code generation, we explore the impact of advanced prompt engineering techniques such as Chain of Thought (CoT)~\cite{wei2022chain} and Retrieval-Augmented Generation (RAG)~\cite{lewis2020retrieval}, providing different insights for optimizing LLM-assisted quantum code generation. Furthermore, by leveraging the domain-specific optimization potential in quantum programming, our framework incorporates a quantum error predictor to facilitate automatic quantum error correction, ensuring robust and fault-tolerant quantum code generation.
Overall, this work makes the following contributions:

\begin{itemize} 
    \item We propose a novel multi-agent framework comprising three key agents: a code generation agent, a semantic analyzer, and a quantum error predictor. The framework employs a multi-pass inference strategy for iterative optimization (Section \ref{framework}). 
    \item We investigate the effectiveness of different advanced prompt engineering techniques, presenting detailed experimental results that provide insights into optimizing LLM-assisted quantum code generation (Section \ref{optimisations}). 
    \item Our framework integrates automatic quantum error correction by leveraging Quantum Error Correction (QEC) decoders, significantly reducing noise in quantum environments (Section \ref{results}). 
\end{itemize}

%% file: 2_background.tex
\section{Background and Related Work}

\subsection{LLMs and Multi-Agent Frameworks} 

Large Language Models (LLMs) are built upon the transformer model \cite{vaswani2017attention}. The most powerful and recent models, such as GPT-4~\cite{openai2023gpt4}, are built upon the decoder-only transformer architecture, in which the output sequence is based on the previous tokens, rather than having separate encoders and decoders. Such models are used for a variety of Natural Language Processing (NLP) tasks, proving to be versatile tools that have the potential to have wide-ranging effects.

One developing area of LLM research is multi-agent frameworks. In this, multiple LLM agents are combined together to allow interaction between different LLMs, which could result in a more accurate response or allow users to use multiple input formats \cite{huang2023agentcoder} \cite{talebirad2023multi}. This has many potential uses, such as safeguarding text generation or a hierarchical chat.

\subsection{Chain-of-Thought and Retrieval-Augmented Generation}

Inference-time optimisation techniques have been shown to improve the quality of output generated by LLMs \cite{zhou2024survey}. These techniques often either augment the prompt with additional information or get the model to think in different ways to better their understanding. Two promising methods are Retrieval-Augmented Generation (RAG) \cite{lewis2020retrieval} and Chain-of-Thought prompting \cite{wei2022chain}. 

RAG involves collecting data or documents relevant to the generation task and building a vectorstore database. We then retrieve the appropriate chunks of data based on a ranking algorithm and augment the prompt with this additional data \cite{lewis2020retrieval}. This gives the model more specific context from which to infer the answer, improving generation accuracy. Chain-of-Thought (CoT) prompting \cite{wei2022chain} encourages models to use logical reasoning when generating answers. Zero-shot CoT \cite{kojima2022large}, in which the model is explicitly told to think "step by step", encourages the usage of logical reasoning. Another technique is manual CoT \cite{wei2022chain} in which a sample question and answer pair is provided to demonstrate the type of reasoning the model should use. Overall, both RAG and CoT have been shown to improve the accuracy of models at the inference stage by augmenting the context or eliciting the logical reasoning ability of the LLM.

\subsection{Quantum Computing}

Quantum computing is a rapidly growing field that shows profound potential. It uses quantum phenomena such as entanglement and superposition to allow information to be stored in an exponentially more dense manner using qubits. Whilst this information cannot be accessed in a one-to-one manner, it does allow access to unique phenomena such as interference and true parallelism, creating new algorithmic possibilities. The ability to exploit these phenomena has allowed quantum algorithms to be made for many problems that are NP or NP-complete in the classical environment. One example is Shor's algorithm for finding prime factors \cite{shor1999polynomial}, which is a problem that is NP for classical computers. This algorithm uses superposition and interference to manipulate qubits in such a way as to find the factors in polynomial time. There are many more examples of algorithms in cryptography \cite{gerjuoy2005shor} and beyond, making quantum computing a field that has exciting possibilities.

One unique challenge posed by developing quantum algorithms is the presence of quantum noise. This is caused by factors such as thermal fluctuations and can result in qubit measurements being incorrect, meaning that experimental results often differ from theoretical results. Much work has been done in the field of quantum error correction (QEC), with many QEC codes being developed, such as the Steane code \cite{steane1996simple}. One common subset of QEC codes are surface codes \cite{fowler2012surface}, which encode one logical qubit onto a lattice of physical qubits. The nature of these codes means that they are topology-dependent, making it difficult to apply QEC codes in a practical development environment.

\subsection{Related Work}

LLM development is undergoing rapid improvement in the field of code generation \cite{jiang2024survey, gai2025exploring}. Models have shown promising ability to capture both the syntax and semantics of a language, with popular models such as CodeLlama \cite{roziere2023code} being able to produce high quality code. These models achieve consistently high scores on coding benchmarks like HumanEval \cite{chen2021evaluating}, which is developed to evaluate models' ability to produce code to the same standard as a human software engineer. Multi-agent frameworks have also been utilized to create a fully-fledged code development environment, with a framework that can produce code and generate an appropriate test suite \cite{huang2023agentcoder}.

Writing code for quantum computers poses significant challenges, as quantum algorithms must be reversible and error tolerant, which demands developers to approach problems in a different way. This creates a unique opportunity for LLMs to aid developers in code writing. Despite the accessibility of the Qiskit library \cite{dupuis2024qiskit}, there is a steep learning curve for beginners to develop novel quantum algorithms. LLMs have the potential to enable developers to create accurate, fault-tolerant code without being experts in the field. IBM's Qiskit Code Assistant model \cite{dupuis2024qiskit} attempts to provide this by fine-tuning on a corpus of Qiskit programs. It achieves better accuracy on a HumanEval style benchmark \cite{vishwakarma2024qiskit} than traditional code generation models but still only achieves 46\% accuracy, suggesting further development is needed. Furthermore, it does not take into account the presence of quantum noise in its code generation, so it will not produce fault-tolerant code.

%% file: 3_framework.tex
\section{Multi-Agent Framework and Training} \label{framework}

\subsection{Framework Overview}

\begin{figure}[t]
    \centering
    \includegraphics[width=0.85\linewidth]{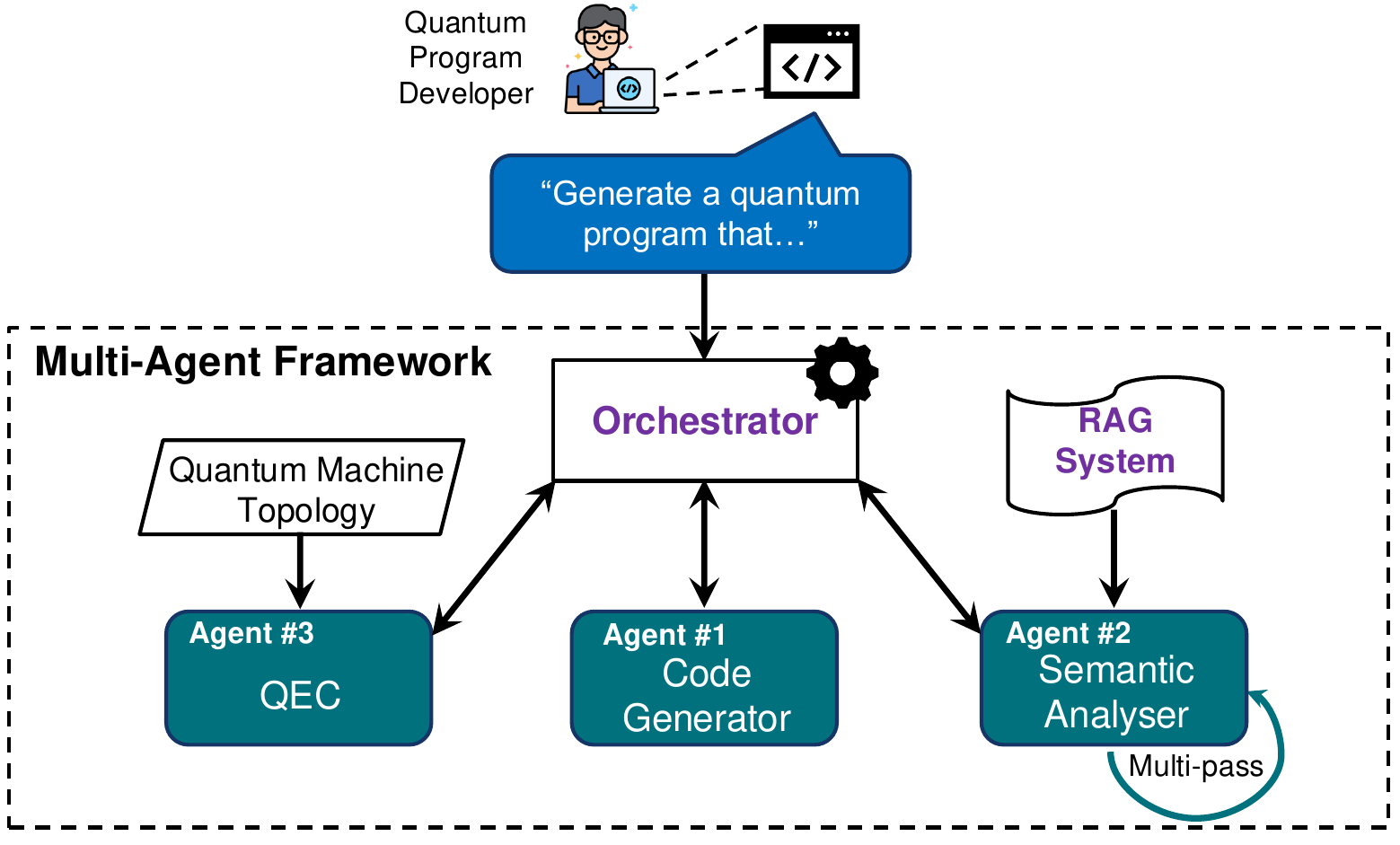}
    \caption{Multi-Agent Design for Quantum Code Generation.}
    \label{fig:1}
\end{figure}

An overview of the proposed multi-agent framework is depicted in~\figref{fig:1}. The framework mainly comprises three agents, each responsible for a specific task: code generation, semantic analysis, and quantum error correction.

\begin{itemize}[leftmargin=*]
    \item \textbf{Code Generation Agent}: We adopt Starcoder~\cite{lozhkov2024starcoder} as the base model for quantum code generation. To enhance the quality of the generated quantum program, we apply supervised fine-tuning on Starcoder using a custom dataset containing the latest open-source Qiskit \cite{dupuis2024qiskit} code. Additionally, we introduce several optimization techniques aimed to further improve the accuracy of the code generation agent, which will be detailed in~\secref{optimisations}.
    \item \textbf{Semantic Analysis Agent}: To enhance the quantum program generated by the initial model with better semantic accuracy, we design a dedicated agent for semantic analysis. This agent can help the code generation process to mitigate the lack of knowledge available in online sources about quantum algorithms, allowing users to incrementally improve the generated code to meet their requirements.
    \item \textbf{QEC Decoder Generation Agent}:
    Given the error-prone nature of quantum computing, primarily caused by quantum noise, it is essential to incorporate Quantum Error Correction (QEC) into our code generation process to enhance the reliability and robustness of the generated quantum programs. 
    To achieve this, we adopt a QEC code decoder to predict and reduce quantum error, allowing our framework to generate fault-tolerant code.
\end{itemize}

By designing the multi-agent framework shown in~\figref{fig:1}, we are able to address the three main challenges in optimizing quantum code generation. It allows us to not only generate syntactically and semantically valid quantum code, but also ensure that we can run the generated code in a stable environment, by the support of AI-generated Error Correction Codes.

\subsection{Training Dataset and Test Suite}

We collected our training data by scraping Github repositories with an open-source license. Due to the fast-paced nature of quantum computing, we filtered the repositories to those updated after February 2024. We found that even filtering by a date this recent still resulted in out-of-date code, including code from official Qiskit Community repositories~\cite{qiskit_machine_learning}. We further filtered the retrieved Python and Jupyter Notebook files based on whether they contained a Qiskit import statement. The notebooks were then split into code and markdown tiles based on sentinel tokens~\cite{dupuis2024qiskit}. After filtering, our total number of tokens was 3M, which was upsampled to 9M, with official sources given higher priority. Whilst this is a small dataset, it did not affect our approach to obtaining meaningful results, as we seek to compare the effectiveness of both fine-tuning techniques and a multi-agent structure. By demonstrating that these optimization techniques and our framework provide a meaningful improvement on the generated quantum code, we can justify working on creating a more accurate code generation agent by creating a larger dataset with higher data quality.

In order to perform our experiments, we created a set of prompt-answer pairs to be passed into our multi-agent system. These prompts cover a wide range of code-generation topics. 
\begin{itemize}[leftmargin=*]
    \item \textbf{Basic}: These tests covers basic code generation for the Qiskit library, allowing us to test the basic syntactic requirements of our model. The tests include basic circuit generation and simulation on quantum devices, ensuring the model can generate and run code on real-world devices.
    \item \textbf{Intermediate}: These cover more complicated quantum circuits and algorithms, with a focus on well-known algorithms, such as Shor's \cite{shor1999polynomial} and Grover's \cite{grover1996fast} algorithms. This is intended to test advanced syntactic code generation, and allow for semantic testing on common applications.
    \item \textbf{Advanced}: This part covers more advanced topics specific to quantum computing. For example, this section contains prompts covering quantum teleportation \cite{bouwmeester1997experimental}, the quantum walk algorithm \cite{santha2008quantum} and quantum annealing \cite{finnila1994quantum}. We expect the model to have little to no knowledge of these algorithms from the base training, so this section can be used to carry out advanced semantic testing.
\end{itemize}
Our test suite consists of 47\% basic tests, 24\% intermediate and 29\% advanced. Compared to the Qiskit HumanEval benchmark \cite{vishwakarma2024qiskit}, our test suite contains much more challenging prompts, asking the model to reason about complex quantum algorithms that the base model would have no knowledge of. This allows us to test how our model performs in advanced settings but may lead to worse performance than QHE.

%% file: 4_optimization.tex
\begin{figure}
    \centering
    \includegraphics[width=0.99\linewidth]{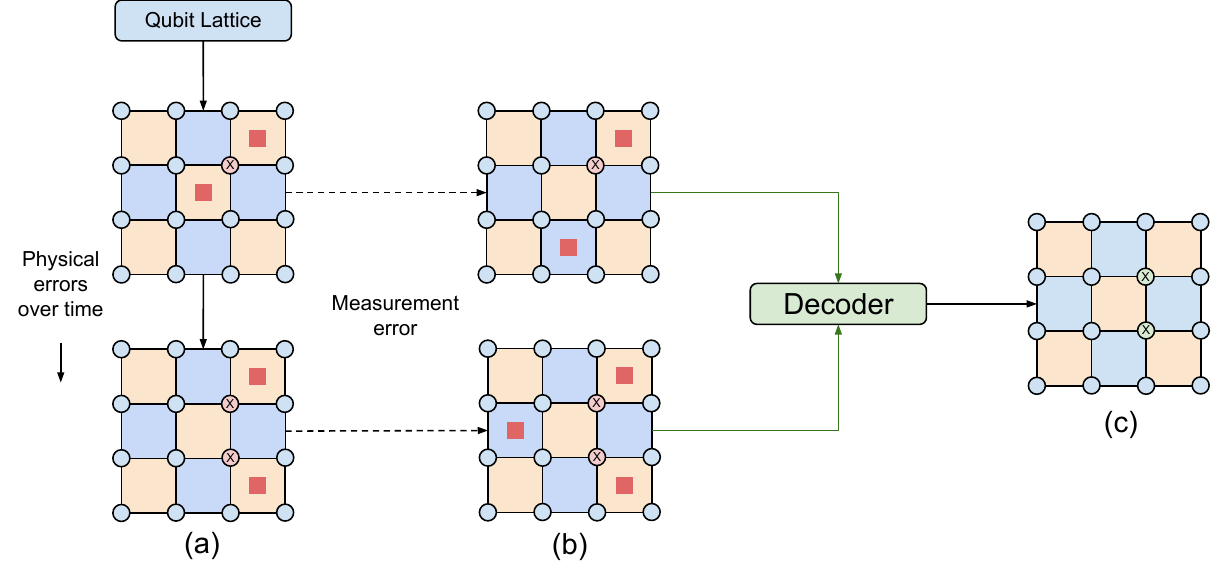}
    \caption{Evolution of qubits during QEC generation. Physical qubits are subject to depolarising noise over time. In (a), X bit-flips violate the yellow (X parity) stabilisers in the surface code syndrome (with blue checking Z parity). We also incur noise when we measure the syndromes in (b). We pass multiple faulty syndromes into the decoder to get the required set of corrections (c). The errors shown are from a circuit preparing the state 1-qubit state $\ket{1}$, discussed in~\secref{qec_exp}.}
    \label{fig:2}
\end{figure}

\section{Optimization} \label{optimisations}

\subsection{Iterative Multi-Pass Optimization}

Due to the limited availability of high-quality training data for quantum programming, the quantum programs generated by the fine-tuned LLM model remain error-prone.
Inspired by recent advances in OpenAI’s GPT-o1, we adopt an inference-time optimization approach to generate high-quality quantum programs. To this end, we implement an iterative multi-pass quantum code generation strategy. Specifically, 
we created a prompt template for multi-pass inference that includes the original prompt, as well as the generated code and error trace. This new prompt is then passed into the model in the hope of fixing the error generated on the previous pass. 

By using a multi-pass structure and iterating upon the previous code, we are able to solve errors as they occur. Since we can pass the incorrect code back into the model multiple times, we can also account for multiple errors, allowing the model to focus on fixing a small, singular error, rather than regenerating the entire program.

\subsection{Quantum Error Correction Enhancement}

The quantum programs generated after passing through the code generation agents and semantic analyzer do not contain any form of quantum error correction.
As a result, running these codes on a physical quantum computer would contain high levels of noise, which reduces the reliability of the generated quantum programs. 
Therefore, it is vital to incorporate the mechanism of error correction to mitigate the potential of quantum error. 

To do this, we include a third agent in the framework, which generates a decoder for a Quantum Error Correction (QEC) surface code~\cite{fowler2012surface, sweke2021reinforcement}. 
As shown in~\figref{fig:2}, this agent uses the topology of the quantum device to generate a decoder that allows a surface error correction code to be used when running the algorithm. This is applied after the code has been generated and does not alter its semantics, only applying a fixed set of operations on the physical qubits immediately before measurement. By applying this code, we extend the average qubit lifetime, which means we are less likely to see qubits affected by quantum noise. This, in turn, reduces the amount of error present in the results, improving their accuracy. 

The main drawback of using a surface code is that they are topology-specific. This results in the model needing to be re-trained every time you want to run the code on a different device, which is extremely inefficient, so finding a way to make a QEC surface code that is topology-agnostic is vital to improving the framework's efficiency.

\subsection{Prompt Engineering Optimisations}

We adopted several techniques to mitigate the unique challenges posed by generating quantum code. The first was Retrieval-Augmented Generation (RAG) \cite{lewis2020retrieval}, to provide the model with knowledge of both the structure of quantum algorithms and the structure of the Qiskit library. We then implemented Chain of Thought (CoT) \cite{wei2022chain} and Structure Chain-of-Thought (SCoT) \cite{li2023structured} prompting, as these have been shown to improve the accuracy of code generation. Our goal for these techniques was to improve the semantic accuracy of the code generated as we found that the lack of algorithmic knowledge available often led to syntactically correct but nonsensical code being generated. The final technique we implemented was multi-pass inference, which was intended to improve the syntactic accuracy of the code.

For our CoT and SCoT prompts, we manually created the first 5 prompts from our testing set using the same techniques demonstrated in previous code generation work \cite{li2023structured}. For each subsequent prompt, we used our examples to generate prompts of the same CoT format using the GPT-4o model \cite{openai2023gpt4}. 

We collated two different RAG datasets. The first scraped the official Qiskit documentation repository on Github, taking the documentation for the latest Qiskit version, including Qiskit-adjacent libraries such as \textit{qiskit-ibm-runtime}. The goal of this dataset is to improve the model's understanding of the library structure, reducing the amount of import errors due to deprecated features. The second consisted of a collection of guides and tutorials explaining the ideas behind and structures of a collection of quantum algorithms. The goal of this dataset is to improve the model's knowledge of algorithms, improving its semantic accuracy. We used the langchain \cite{Chase_LangChain_2022} and ragatouille \cite{clavie2024ragatouille} libraries to create the augmented prompts.

%% file: 5_experiments.tex
\section{Experiments}

\subsection{Experimental Setup}

We chose the StarCoder models~\cite{lozhkov2024starcoder} for fine-tuning since they were pre-trained on a wide corpus of languages and library files, making them suitable for adapting to new programming languages~\cite{li2023starcoder}. 
Furthermore, StarCoder models have demonstrated strong performance across a variety of coding tasks, showing their versatility and suitability for learning in new coding environments~\cite{jiang2024survey}.
In our experiments, we selected StarCoder-3B as the main model for evaluation.

For fine-tuning, we used the \textit{transformers} library \cite{wolf-etal-2020-transformers} with LoRA adapter~\cite{hu2021lora}. To facilitate the use of our custom dataset, we converted it into multiple chunked sets, with randomly applied Fill-in-the-Middle (FIM) transformations \cite{bavarian2022efficient}. 
The chunk size was calculated upon training based on the amount of data provided. We found that the optimal FIM rate (the percentage of chunks to which a FIM transformation is applied) was 0.1. 
Training was conducted for 1500 steps with a batch size of 4. 
The learning rate was linearly increased from 0 to $3 \times 10^{-4}$ over the first 100 warm-up steps and subsequently decayed using a cosine scheduler.

When testing our methods using our testing suite, each optimization method was evaluated independently. 
The CoT and SCoT prompts were passed directly to the model, upon which it generated the code. 
To further evaluate the accuracy of the fine-tuned model without optimizations, we used the \textit{transformers} library~\cite{wolf-etal-2020-transformers} to create a sample of results for each prompt and applied the pass@k metric \cite{chen2021evaluating} to obtain our results.
Additionally, we compared our model with the recent IBM Qiskit Assistant \cite{dupuis2024qiskit} model on the Qiskit HumanEval dataset \cite{vishwakarma2024qiskit}.

\subsection{Overall Accuracy Results}\label{results}

\begin{figure}
    \centering
    \includegraphics[width=0.8\linewidth]{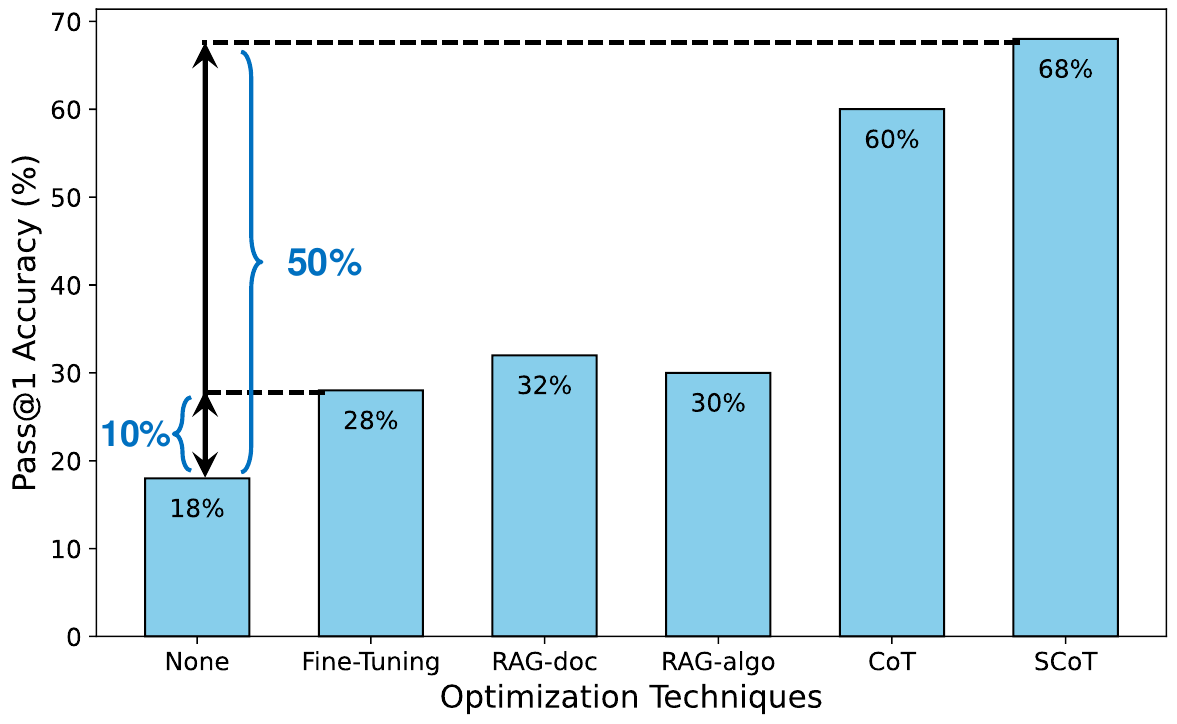}
    \caption{The percentage of results that were semantically and syntactically valid for each technique.}
    \label{fig:accuracy_comparison_fine_tuning_methods}
\end{figure}

\begin{table}
    \begin{tabularx}{0.5\textwidth} { 
  | >{\raggedright\arraybackslash}X 
  | >{\centering\arraybackslash}X 
  | >{\raggedleft\arraybackslash}X | }
 \hline
 Model & QHE Score  \\
 \hline
 Starcoder2-7B  & 17.9\%  \\
\hline
 Starcoder2-7B-QK & 24.5\% \\
\hline
 Starcoder2-7B-QKRAG & 33.8\% \\
\hline
 Starcoder2-7B-QKCoT & 41.4\% \\
\hline
 IBM Granite-20B-CODE-QK & 46.5\% \\
\hline
\end{tabularx}
    \caption{Qiskit HumanEval performance.}
    \label{tab:4}
\end{table}

\begin{figure*}
    \centering
    \includegraphics[width=0.85\linewidth]{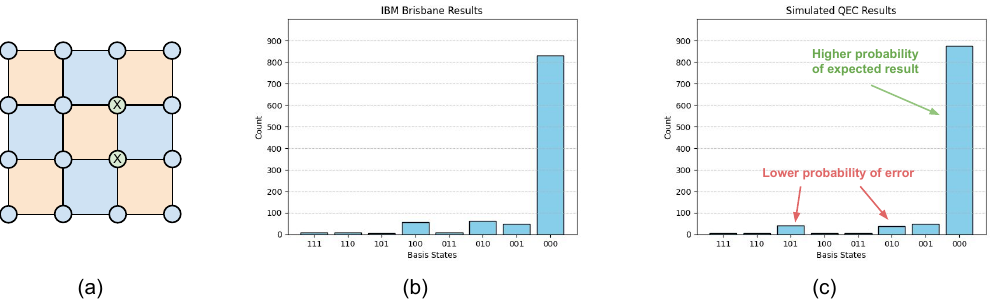}
    \caption{Results for QEC Experiments. (a) shows the corrections suggested by the decoder, (b) the results from running on IBM Brisbane and (c) the new results after applying the corrections.}    \label{fig:grover}
\end{figure*}

Using our test suite, we were able to evaluate the accuracy of applying our multi-agent framework with the optimization techniques proposed.
The overall results of the experiments are shown in~\figref{fig:accuracy_comparison_fine_tuning_methods}. 
The results provided are the percentage of prompts that are both syntactically and semantically correct. By training on the dataset of Qiskit repositories, we were able to increase the  pass@1 metric by $10\%$, up to $28\%$ overall.
\tabref{tab:4} presents the accuracy of the model on the Qiskit HumanEval dataset. 
We observe that both RAG and CoT greatly increase the performance of our 7B model, with CoT prompting achieving results approaching those of the 20B IBM Qiskit model. 
Although there is a performance gap of approximately $5\%$, we plan to adopt a more powerful base model in future evaluations.

\subsection{Effects of RAG, CoT, and SCoT}

As shown in~\figref{fig:accuracy_comparison_fine_tuning_methods}, applying RAG~\cite{lewis2020retrieval} optimization had negligible impact on the overall accuracy of the model, likely because it had little effect on semantic accuracy. The improvement on the Qiskit HumanEval benchmark mainly came from resolving syntactic errors. We used a basic RAG splitting technique, which does not take into account code structure, so we could see better accuracy if we used a more intelligent method.
The results for CoT~\cite{wei2022chain} and SCoT~\cite{li2023structured} show remarkable increases, with CoT leading to an increase of $32\%$ and SCoT an increase of $40\%$ from the fine-tuned model. 
The majority of this increase came from a better semantic understanding of how the algorithms are structured, suggesting that the initial training dataset did not contain many high-quality algorithmic examples. 

We also noticed that CoT prompting led to a decrease in the number of results that were syntactically valid but semantically incorrect in the Qiskit HumanEval dataset \cite{vishwakarma2024qiskit}. For the RAG model, we saw 45.7\% syntactic accuracy, but only 33.8\% semantic accuracy. For CoT, we achieved a similar 46.4\% accuracy but now achieved 41.4\% semantic accuracy, showing that CoT greatly improved semantic understanding.
We are able to achieve higher accuracy on our test suite compared to Qiskit HumanEval. This is likely because we are testing semantic knowledge, rather than Qiskit specific syntax, so CoT prompting allows us to get high accuracy.

\subsection{Effects of Multi-Pass Inference and QEC} \label{qec_exp}

We found that applying multi-pass inference into the semantic analyzer model can improve the accuracy to 34\% using triple passes. However, additional inference passes, despite incurring higher computational costs, yielded limited benefit. The diminishing accuracy improvement is likely due to the nature of the error, which was mostly the misuse of imports or the use of deprecated code. 
If these issues were resolved, the results did demonstrate evidence that multi-pass inference could be used to improve the model's ability to resolve other syntactic and semantic errors. But note that to improve accuracy further, we first need to solve the limited dataset problem, as we require extra prompt-error-answer examples to further train this feature.

Leveraging our multi-agent framework with the support of quantum error prediction and correction, we were able to reduce the amount of quantum error in our experiment results. 
\figref{fig:grover} shows an example of the constant Deutsch-Jozsa oracle~\cite{deutsch1992rapid} under a quantum noise environment, with and without the use of our framework. We expect the circuit to yield the $\ket{000}$ state, as we pass in a constant function to the oracle.
By applying the corrections suggested by the decoder, we increase the average qubit lifetime, decreasing the probability of an erronenous measurement. Due to the fact that we cannot directly alter physical qubits on IBM devices with corrections, we simulated our results for (c) using a lower error probability than IBM Brisbane \cite{brisbane2025}, corresponding to the new error rate after QEC.
Although this example with surface code applied is topology-dependent, it is sufficient to demonstrate the effectiveness of our framework in reducing quantum error.

\subsection{Observation and Future Directions} 

Different from LLM-assisted code generation for general-purpose programming languages, our experimental results reveal that different optimization techniques have vastly distinct effects on quantum code generation.
Applying RAG enhancement resulted in only a marginal semantic accuracy increase, likely attributable to the fact that the documentation available for Qiskit is not up to date, preventing the effective mitigation of import errors.
In contrast, other techniques such as Chain-of-Thought (CoT) reasoning introduced significant accuracy improvement. 
This improvement is attributed to the enhanced semantic knowledge provided to the model, allowing it to correctly structure the generated quantum program. 
This approach likely outperformed RAG as it allowed us to more directly inform the model’s decision-making process, rather than inferring how the algorithms work from the, rather limited, dataset we provided. It is also important to note that some of the errors occur due to incorrect CoT prompt generation, meaning the model generated syntactically correct but semantically invalid code. So, a necessary future step may be improving the model used for CoT generation.

One limitation encountered during our experiments was the limited dataset sizes, which can be evidenced by both the poor increase in the fine-tuned models' pass@1 accuracy and the lack of improvement from applying the RAG technique. 
An important direction of our future work is to collect a larger and higher-quality dataset. This presents challenges due to the lack of data online and the rapidly changing nature of the field. 
One potential solution is to scrape other sites, such as the official Qiskit website and forums, while supplementing the dataset with hand-built examples.
Other promising directions include the generation of synthetic data~\cite{yang2024synthetic}, federated learning~\cite{zhao2024clues, zhao2024breaking} and inference-time optimizations~\cite{snell2024scaling, chen2024hardware, chen2024progressive}.

Moreover, a key area for future development is creating a topology-agnostic QEC decoder model. 
Currently, our work includes a model that is topology-specific and requires the devices to follow a fully-connected lattice design, requiring retraining each time algorithms are adapted for different quantum computers.
Whilst our work has shown that using a model like this can aid in producing fault-tolerant code, having a topology-specific model is not a scalable method. 
Developing a model capable of generating quantum error correction decoders for arbitrary topologies would significantly enhance the scalability of fault-tolerant quantum code generation.

%% file: 6_conclusion.tex
\section{Conclusion}

This paper proposes a novel multi-agent framework to provide an accurate, fault-tolerant quantum computing code generation using Large Language Models (LLMs). We investigate the effect of various optimizations, including Chain-of-Thought, Retrieval-Augmented Generation and multi-pass inference.
Our experiments highlight the need for multi-agent design, as current LLMs lack deep algorithmic understanding.
Incorporating a Quantum Error Correction agent, we observe a reduction in average noise on quantum hardware.
Future work includes enhancing the generality of our methods and exploring FPGA-based acceleration for the agent-based system.